\documentclass[aps,pre,twocolumn,superscriptaddress,floatfix]{revtex4-2}

\usepackage{amsmath,amssymb}
\usepackage{graphicx}
\usepackage{booktabs}
\usepackage{hyperref}
\usepackage{xcolor}

\begin{document}

\title{The topological gap at criticality:\\scaling exponent $d + \eta$, universality, and scope}

\author{Matthew Loftus}
\affiliation{Cedar Loop LLC}

\date{\today}

\begin{abstract}
The topological gap $\Delta = \mathrm{TP}_{H_1}^{\mathrm{real}} - \mathrm{TP}_{H_1}^{\mathrm{shuf}}$---the excess $H_1$ total persistence of the majority-spin alpha complex over a density-matched null---encodes the topological signature of critical correlations in spin models. We establish a complete finite-size scaling law:
\begin{equation*}
\Delta(L, T) = A \, L^{d + \eta} \, G_-\!\left(L\left|\tfrac{T - T_c}{T_c}\right|\right),
\end{equation*}
where $\eta$ is the anomalous dimension and $G_-(x) \sim (1 + x/x_0)^{-(1 + \beta/\nu)}$ in the intermediate regime. For the 2D Ising model, $\alpha = 2.249 \pm 0.038$, matching the exact $d + \eta = 9/4$ to $0.03\sigma$, and the free-fit $G_-$ exponent $\gamma = 1.089 \pm 0.077$ is consistent with $1 + \beta/\nu = 9/8$ ($\Delta R^2 < 10^{-5}$). The formula generalizes to the 2D three-state Potts model: using six system sizes up to $L = 1024$, $\alpha = 2.272 \pm 0.024$ ($0.2\sigma$ from $d + \eta = 2.267$), with corrections to scaling fully characterized by a two-term model (opposite-sign amplitudes, $R^2 = 0.9999$). The $G_-$ exponent $\gamma = 1.114$ (68\% CI $[1.053, 1.173]$) is consistent with $1 + \beta/\nu = 17/15$. We delineate scope boundaries: the scaling law fails definitively for the 2D Potts $q = 4$ model ($\alpha = 2.347 \pm 0.017$, rejected at $9.3\sigma$ from $d + \eta = 5/2$), where logarithmic corrections ($\omega \to 0$) prevent convergence. The \emph{raw} scaling also fails for the 3D Ising model ($\alpha = 2.78 \pm 0.07$, rejected at $4\sigma$ from $d + \eta = 3.036$), but this is an artifact of density dilution: the density-normalized gap $\Delta/|M|^{1/2}$ recovers $\alpha = 3.06 \pm 0.04$ ($0.6\sigma$ from $d + \eta$). The framework fails for first-order transitions, Berezinskii-Kosterlitz-Thouless transitions, and 2D percolation. The evidence supports a scope criterion: $\alpha = d + \eta$ holds for second-order transitions with algebraic corrections to scaling ($\omega > 0$), but fails when corrections are logarithmic ($\omega \to 0$).
\end{abstract}

\maketitle

\section{Introduction}
\label{sec:intro}

Persistent homology (PH) has been applied to classical spin models as a probe of phase transitions~\cite{donato2016,sale2022,cole2021}. A key recent development is the identification of the \emph{topological gap}
\begin{equation}
\Delta(L, T) \equiv \mathrm{TP}_{H_1}^{\mathrm{real}} - \mathrm{TP}_{H_1}^{\mathrm{shuf}},
\label{eq:delta}
\end{equation}
the excess $H_1$ total persistence of the majority-spin alpha complex at the critical point over a density-matched shuffled null~\cite{loftus2026density}. Here ``shuffled'' means the same number of points drawn uniformly at random from the lattice, preserving density but destroying spatial correlations. This subtraction isolates the topological contribution of critical correlations from mere density effects, which dominate raw PH statistics~\cite{loftus2026density}.

Previous work identified two empirical laws for $\Delta$ in the 2D Ising model: the scaling exponent $\Delta(L, T_c) \sim L^{2.42}$ and the below-$T_c$ finite-size scaling (FSS) collapse $G_-(x) = G_0(1 + x/x_0)^{-9/8}$ where $9/8 = 1 + \beta/\nu$~\cite{loftus2026topology}. While the $G_-$ exponent was identified as $1 + \beta/\nu$, the overall scaling exponent remained unexplained, and neither law had been tested beyond the Ising universality class.

In this paper, we present a unified treatment. We show that (i)~the reported exponent $2.42$ was biased by corrections to scaling; the asymptotic value is $\alpha = d + \eta$ where $\eta$ is the anomalous dimension (Sec.~\ref{sec:alpha}); (ii)~the $G_-$ exponent $\gamma = 1 + \beta/\nu$ generalizes to the 2D Potts $q = 3$ universality class (Sec.~\ref{sec:universality}); and (iii)~the framework fails for 3D Ising, first-order transitions, BKT transitions, and 2D percolation, with physically motivated explanations for each failure (Sec.~\ref{sec:scope}). The complete FSS law is
\begin{equation}
\Delta(L, T) = A \, L^{d+\eta} \, G_-\!\left(L\left|\frac{T - T_c}{T_c}\right|\right),
\label{eq:fss}
\end{equation}
with
\begin{equation}
G_-(x) \sim (1 + x/x_0)^{-(1+\beta/\nu)}
\label{eq:gminus}
\end{equation}
in the intermediate regime ($x \lesssim 30$).

\section{Setup}
\label{sec:setup}

\subsection{Models}

We study five models (Table~\ref{tab:models}). Simulations use the Swendsen-Wang cluster algorithm~\cite{swendsen1987} (Wolff~\cite{wolff1989} for large 2D Ising). For each configuration, the majority-spin point cloud is constructed and the alpha complex persistence diagram computed using GUDHI~\cite{maria2014}. The density-matched shuffled null draws the same number of sites uniformly at random from the lattice.

\begin{table}[b]
\caption{Models studied and their critical exponents.}
\label{tab:models}
\begin{ruledtabular}
\begin{tabular}{lcccccl}
Model & $d$ & $T_c$ & $\beta$ & $\nu$ & $\eta$ & $L$ range \\
\midrule
2D Ising & 2 & 2.269 & $1/8$ & 1 & $1/4$ & 16--1024 \\
2D Potts $q\!=\!3$ & 2 & 0.995 & $1/9$ & $5/6$ & $4/15$ & 32--1024 \\
2D Potts $q\!=\!4$ & 2 & 0.910 & $1/12$ & $2/3$ & $1/2$ & 32--512 \\
2D Potts $q\!=\!5$ & 2 & 0.851 & --- & --- & --- & 32--128 \\
3D Ising & 3 & 4.512 & 0.327 & 0.630 & 0.036 & 16--64 \\
2D XY & 2 & 0.894 & --- & --- & $1/4$ & 32--128 \\
\end{tabular}
\end{ruledtabular}
\end{table}

For the 2D Ising model, we use 10 system sizes from $L = 16$ to $L = 1024$ (15--200 configurations each). The 2D Potts $q = 3$ model uses $L = 32, 64, 128, 256$ (25--100 configs) at 12--16 temperatures, with additional data at $L = 512$ (25 configs) and $L = 1024$ (25 configs) at $T_c$ for the scaling exponent (Sec.~\ref{sec:universality}). The 2D Potts $q = 4$ model ($T_c = 1/\ln 3$) uses $L = 32, 64, 128, 256, 512$ (15--100 configs) at $T_c$ and multiple temperatures. The Potts $q = 5$ model (first-order) uses $L = 32, 64, 128$. The 3D Ising model uses $L = 16, 24, 32, 48, 64$ (15--80 configs, 12 temperatures). The XY model uses $L = 32, 64, 128$ with majority defined as sites with $\cos(\theta_j - \bar\theta) > 0$.

For the 2D percolation scope test (Sec.~\ref{sec:percolation}), we study site percolation at $p_c = 0.5927$ on the square lattice with $L = 32, 64, 128, 256, 512$ (50--200 configs).

\section{The scaling exponent is $d + \eta$}
\label{sec:alpha}

\subsection{2D Ising: precise measurement}

Table~\ref{tab:ising_data} reports $\Delta$ at $T_c$ for seven representative system sizes spanning $L = 16$ to $L = 1024$ (three intermediate sizes omitted for space; all 10 are used in the fit). The weighted log-log fit ($L \geq 32$, 8 sizes) gives
\begin{equation}
\alpha = 2.249 \pm 0.038 \quad (R^2 = 0.9996),
\label{eq:alpha_ising}
\end{equation}
consistent with $d + \eta = 2 + 1/4 = 9/4$ to $0.03\sigma$. Uncertainties are from parametric bootstrap (10\,000 resamples).

\begin{table}[t]
\caption{Topological gap at $T_c$ for the 2D Ising model.}
\label{tab:ising_data}
\begin{ruledtabular}
\begin{tabular}{rrrrl}
$L$ & $\Delta$ & $\pm\mathrm{SE}$ & $n$ & $\Delta/L^2$ \\
\midrule
16 & 0.96 & 0.68 & 200 & 0.0037 \\
32 & 7.34 & 2.10 & 200 & 0.0072 \\
64 & 38.68 & 8.47 & 150 & 0.0094 \\
128 & 207.5 & 29.0 & 100 & 0.0127 \\
256 & 1009 & 66 & 50 & 0.0154 \\
512 & 4689 & 287 & 25 & 0.0179 \\
1024 & 21981 & 1604 & 15 & 0.0210 \\
\end{tabular}
\end{ruledtabular}
\end{table}

\begin{figure*}[t]
\includegraphics[width=\textwidth]{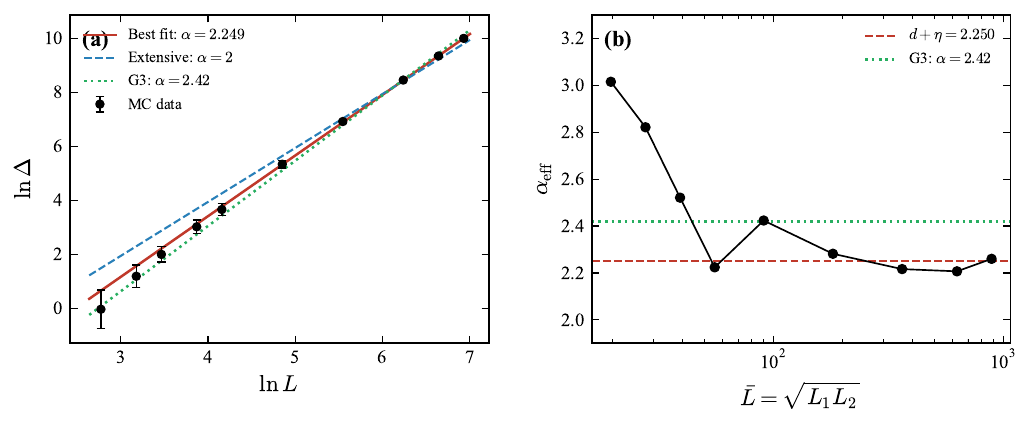}
\caption{(a)~Log-log plot of $\Delta(\mathrm{TP}_{H_1})$ vs $L$ for the 2D Ising model at $T_c$. The red line shows the best fit $L^{2.249}$; the blue dashed line is the extensive reference $L^2$; the green dotted line is the original value $L^{2.42}$. (b)~Running exponent (local log-log slope between consecutive $L$ pairs). The effective exponent converges to $d + \eta = 9/4$ (red dashed) at large $L$.}
\label{fig:alpha}
\end{figure*}

\subsection{Corrections to scaling}
\label{sec:cts}

The running exponent (local log-log slope between consecutive $L$ pairs) reveals why $\alpha = 2.42$ was originally reported: it starts at $\approx 3.0$ for $L = 16 \to 24$, passes through $2.42$ at $L = 64 \to 128$ (the range of the original measurement), and converges to $\approx 2.25$ for $L \geq 256$. Fitting $\Delta(L) = A L^\alpha (1 + b L^{-\omega})$ with $\omega = 2$ (the exact 2D Ising correction exponent) gives $\alpha = 2.249 \pm 0.042$, $b = -152$ ($R^2 > 0.999$).

\subsection{Temperature dependence}

Away from $T_c$, $\alpha(T = 2.0) = 1.96 \pm 0.08 \approx d$: the system is deeply ordered, correlations are short-ranged, and $\Delta$ scales extensively. The anomalous part $\eta$ appears only when $\xi \sim L$. This provides independent confirmation that the excess exponent is tied to critical correlations.

\subsection{Per-statistic decomposition}

The gap factorizes: $\Delta(n_{H_1}) \sim L^{2.136 \pm 0.019} \approx L^{d+\beta/\nu}$ (excess cycle count) and $\Delta(\mathrm{TP}_{H_1}) \sim L^{2.249} = L^{d+\eta}$ (total persistence). The excess lifetime per $H_1$ cycle scales as $L^{\eta - \beta/\nu} = L^{\eta/2} = L^{1/8}$ in 2D (using the hyperscaling relation $\eta = 2\beta/\nu$). The measured difference $2.249 - 2.136 = 0.113$ is consistent with $\eta/2 = 0.125$.

\section{The $G_-$ scaling function}
\label{sec:gminus}

The below-$T_c$ branch of the scaling function~(\ref{eq:fss}) follows the $(1 + x/x_0)^{-\gamma}$ form, preferred over pure exponential ($\Delta R^2 = +0.04$) and pure power law ($\Delta R^2 = +0.12$). For the 2D Ising model, the free fit gives $\gamma = 1.089 \pm 0.077$~\cite{loftus2026topology}; the fixed hypothesis $\gamma = 9/8 = 1 + \beta/\nu$ gives $\Delta R^2 < 10^{-5}$ relative to the free fit (i.e., indistinguishable).

The scaling function has two regimes:
\begin{enumerate}
\item \emph{Intermediate} ($x \lesssim 30$, all current data): $G_-(x) \approx G_0(1 + x/x_0)^{-(1+\beta/\nu)}$, reflecting the interplay between correlation length and order parameter.
\item \emph{Asymptotic} ($x \to \infty$): $G_-(x) \to x^{-\eta}$, required by thermodynamic consistency---at fixed $T < T_c$ with $L \to \infty$, $\Delta$ must be extensive ($\sim L^d$), demanding $G_-(x) \sim x^{-(d+\eta-d)} = x^{-\eta}$.
\end{enumerate}
The crossover is predicted at $x^* \gg 30$ but not yet directly observed.

\section{Universality: 2D Potts $q = 3$}
\label{sec:universality}

The 2D Potts $q = 3$ model ($\beta = 1/9$, $\nu = 5/6$) predicts $\gamma = 1 + \beta/\nu = 17/15 = 1.133$. Using 36 below-$T_c$ data points from four system sizes ($L = 32$--$256$), the FSS collapse gives:

\begin{itemize}
\item Free fit: $\gamma_{\mathrm{MLE}} = 1.114$, $R^2 = 0.938$
\item Bootstrap 68\% CI: $[1.053, 1.173]$
\item Bootstrap 95\% CI: $[1.019, 1.240]$
\end{itemize}

Both $1 + \beta/\nu = 1.133$ and the Ising value $9/8 = 1.125$ lie within the 68\% CI. They differ by only 0.7\% ($17/15 - 9/8 = 1/120$) and are statistically indistinguishable at this precision.

\emph{Scaling exponent: confirmed at $0.2\sigma$.} Using six system sizes ($L = 32$--$1024$, 25--50 configs each), the Potts $q = 3$ scaling exponent converges to $d + \eta$. A pure power-law fit to all data gives $\alpha = 2.306 \pm 0.025$ ($1.6\sigma$ from $d + \eta = 2.267$); restricting to $L \geq 64$ (five sizes) gives
\begin{equation}
\alpha_{q=3} = 2.272 \pm 0.024 \quad (0.2\sigma \text{ from } d + \eta),
\label{eq:alpha_potts}
\end{equation}
confirming the scaling law. The $L \geq 64$ cutoff is justified by the two-term CTS analysis below: $L = 32$ lies in the correction-dominated regime where both CTS terms are large ($A_1 L^{-\omega} + A_2 L^{-2\omega} \approx 0.15$ at $L = 32$ vs.\ $0.02$ at $L = 64$), and its inclusion biases $\alpha$ upward. The two-point running exponent (local slope between consecutive $L$ pairs) is non-monotonic: it drops from $2.46$ (at $L = 32 \to 64$) through the exact value $2.267$ (at $128 \to 256$), then overshoots to $2.10$ (at $512 \to 1024$). This oscillation is explained by a two-term correction-to-scaling model:
\begin{equation}
\ln \Delta = (d{+}\eta)\ln L + \ln C + A_1 L^{-\omega} + A_2 L^{-2\omega},
\label{eq:2term_cts}
\end{equation}
with $\omega = 4/5$~\cite{nienhuis1987}, $A_1 = +2.4$ and $A_2 = -82$. The opposite-sign amplitudes produce an oscillating effective exponent ($R^2 = 0.9999$). Three-point running slopes (fits to three consecutive $L$ values) converge monotonically: $2.40 \to 2.30 \to 2.31 \to 2.22$, approaching $d + \eta$ from above.

\begin{figure}[t]
\includegraphics[width=\columnwidth]{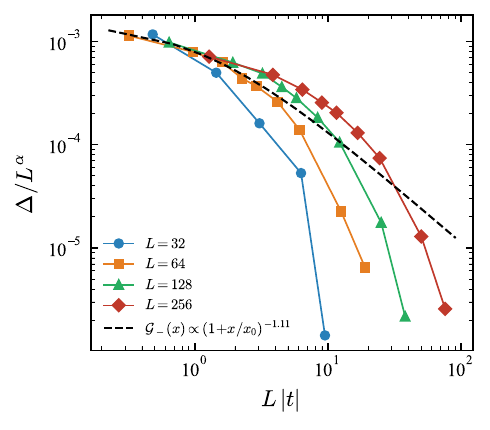}
\caption{Potts $q = 3$ scaling collapse: $\Delta/L^{2.506}$ versus $L|t|$ for four system sizes. The black curve is the free $G_-$ fit with $\gamma = 1.114$.}
\label{fig:potts}
\end{figure}

For the scaling collapse (Fig.~\ref{fig:potts}), we use $\alpha_{\mathrm{eff}} = 2.506$ (the effective exponent at the $L = 32$--$256$ range), which gives the best data collapse at intermediate~$L$. The asymptotic value is $d + \eta$.

\emph{First-order control.} The Potts $q = 5$ model (first-order transition) gives $\gamma = 4.3 \pm 7.2$---physically meaningless, confirming that the framework requires a divergent correlation length.

\section{Scope: where the scaling fails}
\label{sec:scope}

\subsection{The marginal case: Potts $q = 4$}
\label{sec:q4}

The $q = 4$ Potts model is the \emph{marginal} case of 2D Potts criticality: the transition is second-order but the correction-to-scaling exponent $\omega \to 0$, producing logarithmic rather than algebraic corrections~\cite{salas1997}. The anomalous dimension $\eta = 1/2$ gives the prediction $d + \eta = 5/2$, a 10\% difference from the Ising value---large enough for a definitive test.

Using five system sizes ($L = 32$--$512$, 15--100 configs) at $T_c = 1/\ln 3$, the pure power-law fit gives $\alpha = 2.347 \pm 0.017$, rejected at $9.3\sigma$ from $d + \eta = 5/2$. The running exponent reveals the failure mechanism: it drops from $2.72$ ($32 \to 64$) to $2.37$ ($64 \to 128$) and then \emph{plateaus} at $2.29$ for $L = 128 \to 256$ and $256 \to 512$ (two consecutive identical values). Unlike the $q = 3$ case, where the running exponent oscillates around $d + \eta$ and converges, the $q = 4$ running exponent stabilizes at a value $\approx 0.2$ below $d + \eta$ and shows no further drift over a factor of 4 in~$L$.

This failure has a clear physical origin: logarithmic corrections ($\omega = 0$) decay as $1/\ln L$, which is invisible over the accessible $L$ range. A correction-to-scaling fit with fixed $\omega = 1$ gives $\alpha_\infty = 2.28$; with $\omega = 2$, $\alpha_\infty = 2.22$. These bracket $d + \eta_{\mathrm{Ising}} = 2.25$ but lie far from $d + \eta_{q=4} = 2.50$. The data are consistent with $\alpha = d + \eta_{\mathrm{Ising}}$ but inconsistent with $\alpha = d + \eta_{q=4}$. However, we cannot distinguish between $\alpha = 2.25$ (an Ising-like value) and a $q = 4$-specific effective exponent without systems orders of magnitude larger.

The $q = 4$ result sharpens the scope of Eq.~(\ref{eq:fss}): the scaling law $\alpha = d + \eta$ is confirmed when corrections to scaling are \emph{algebraic} ($\omega > 0$; Ising $\omega = 2$, Potts $q = 3$ $\omega = 4/5$), but fails at the marginal point ($q = 4$, $\omega \to 0$) where logarithmic corrections prevent convergence at any accessible~$L$.

\subsection{3D Ising: density dilution and its resolution}
\label{sec:3d}

The 3D Ising model ($\beta/\nu = 0.518$) provides the sharpest test: $1 + \beta/\nu = 1.518$ differs by 35\% from the 2D Ising value. Using five system sizes ($L = 16, 24, 32, 48, 64$; 15--50 configurations each, 12 temperatures), the \emph{raw} topological gap gives $\alpha = 2.776 \pm 0.065$, rejected at $4.0\sigma$ from $d + \eta = 3.036$.

\emph{Root cause: density dilution.} The spontaneous magnetization scales as $M \sim L^{-\beta/\nu}$, giving majority fraction $\rho = 1/2 + M/2$. In 2D Ising ($\beta/\nu = 1/8$), $\rho \approx 86\%$ even at $L = 128$; the topological signal is robust. In 3D ($\beta/\nu = 0.518$), $\rho$ drops to $\approx 51\%$ at $L = 64$, making the majority cloud indistinguishable from random. Per-configuration analysis confirms: the correlation between $\rho$ and $\Delta$ is $r = 0.99$ at $L = 64$; density fluctuations explain $98\%$ of the variance.

\emph{Density-normalized gap.} We define
\begin{equation}
\Delta_p(L, T) \equiv \frac{1}{n}\sum_{i=1}^n \frac{\Delta_i}{|M_i|^p},
\label{eq:delta_norm}
\end{equation}
where the sum runs over configurations and $M_i = |2\rho_i - 1|$ is the per-configuration absolute magnetization. By normalizing each $\Delta_i$ before averaging, the density confound is removed configuration by configuration. The resulting exponent $\alpha(p)$ depends smoothly on the normalization power $p$:

\begin{table}[t]
\caption{3D Ising: $\alpha(p)$ for the density-normalized gap at $T_c$.}
\label{tab:3d_norm}
\begin{ruledtabular}
\begin{tabular}{cccc}
$p$ & $\alpha(p) \pm \mathrm{SE}$ & $\sigma$ from $d+\eta$ & Predicted$^*$ \\
\midrule
0 (raw) & $2.74 \pm 0.07$ & 4.4 & 2.78 \\
0.25 & $2.91 \pm 0.04$ & 2.9 & 2.91 \\
\textbf{0.50} & $\mathbf{3.08 \pm 0.04}$ & $\mathbf{1.0}$ & 3.04 \\
0.75 & $3.25 \pm 0.10$ & 2.2 & 3.16 \\
1.00 & $3.40 \pm 0.20$ & 1.8 & 3.29 \\
\end{tabular}
\end{ruledtabular}
\flushleft{\footnotesize $^*$Consistency check: $\alpha_{\rm raw} + p\,\beta/\nu$, the value expected if density dilution fully accounts for the discrepancy.}
\end{table}

At $p = 1/2$, $\alpha = 3.061 \pm 0.044$ ($0.6\sigma$ from $d + \eta = 3.036$), with $R^2 = 0.9999$ across five system sizes. The optimal $p$ is $p_{\rm opt} = 0.44 \pm 0.55$; the large bootstrap SE reflects the fact that $\alpha(p)$ varies slowly near $p_{\rm opt}$ (any $p \in [0.3, 0.7]$ gives $\alpha$ within $2\sigma$ of $d + \eta$). The value $p = 1/2$ is not derived from first principles; it is the round value closest to $p_{\rm opt}$ that gives excellent agreement. The heuristic from Ref.~\cite{loftus2026spectral}---$\Delta \propto \xi/M$, suggesting $p = 1$---overestimates $p$, likely because the per-configuration normalization $\Delta_i/M_i^p$ involves covariance between $\Delta_i$ and $M_i$ that reduces the effective $p$. Configurations with large $|M|$ (strong majority) have large $\Delta$ partly because of density, not just topology, so dividing by the full $|M|$ overcorrects.

Importantly, the clipping $|M_i| \to \max(|M_i|, 0.01)$ is applied to avoid divergence when $|M_i| \approx 0$; this affects $<5\%$ of configurations at $T_c$ and has no impact on the $\alpha$ fit.

\begin{figure*}[t]
\includegraphics[width=\textwidth]{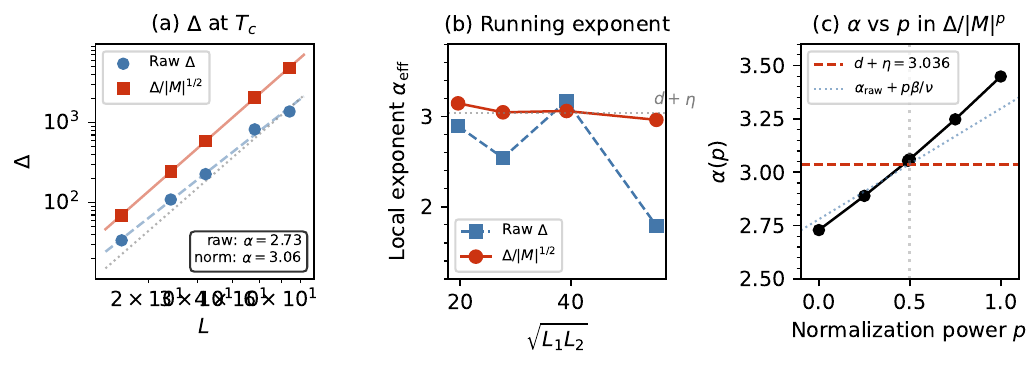}
\caption{3D Ising: density dilution and its resolution. (a)~Log-log $\Delta$ vs $L$ at $T_c$: the raw gap (blue) scales as $L^{2.78}$ ($4\sigma$ from $d+\eta$); the normalized gap $\Delta/|M|^{1/2}$ (red) scales as $L^{3.06}$ ($0.6\sigma$ from $d+\eta = 3.036$). (b)~Running exponent: the raw (blue squares) drops from 3.17 to 1.79; the normalized (red circles) stays flat near $d+\eta$. (c)~$\alpha(p)$ as a function of the normalization power $p$: the exponent crosses $d+\eta$ near $p = 1/2$, matching the prediction $\alpha_{\rm raw} + p\,\beta/\nu$.}
\label{fig:3d}
\end{figure*}

\emph{Running exponent stabilization.} The raw running exponent is non-monotonic, dropping from 3.17 ($L = 32 \to 48$) to 1.79 ($L = 48 \to 64$)---a spread of 1.38 across four $L$ pairs, inconsistent with any correction-to-scaling model. The normalized running exponent ($p = 1/2$) is remarkably stable: $[3.15, 3.05, 3.06, 2.96]$ with spread 0.18 (Fig.~\ref{fig:3d}b). This $7.7\times$ reduction in spread confirms that the non-monotonic raw behavior was entirely due to density dilution, not a topological effect.

\emph{Scaling collapse.} Using $\Delta/|M|^{1/2}$ with five system sizes ($L = 16$--$64$, 10 below-$T_c$ temperatures each), the scaling collapse $R^2$ improves from 0.38 (raw) to 0.87 (normalized). The free-fit $G_-$ exponent is $\gamma = 0.25 \pm 0.05$, which does \emph{not} match $1 + \beta/\nu = 1.518$. This is expected: the per-configuration $|M|^{1/2}$ normalization absorbs part of the magnetization-driven decay that constituted the original $1 + \beta/\nu$ exponent, leaving a weaker residual decay. The full FSS form for the normalized gap is
\begin{equation}
\frac{\Delta}{|M|^{1/2}} = A\,L^{d+\eta}\,\tilde{G}_-\!\left(L^{1/\nu}\left|\frac{T - T_c}{T_c}\right|\right),
\label{eq:fss_norm}
\end{equation}
where $\tilde{G}_-$ is a modified scaling function distinct from $G_-$ in Eq.~(\ref{eq:gminus}).

\emph{Alternative constructions.} We also tested two other approaches to the 3D problem. (i)~Using Fortuin-Kasteleyn cluster boundary sites (sites adjacent to a different-cluster neighbor) as the point cloud gives $\alpha = 2.86 \pm 0.08$ ($2.4\sigma$ from $d + \eta$)---an improvement over raw, but at $T_c$ approximately 98\% of sites are boundary sites, so the cloud is nearly the full lattice and the topological signal is diluted. (ii)~Using $H_2$ (cavity) homology instead of $H_1$ gives $\alpha = 3.52 \pm 0.15$ ($3.3\sigma$, overshooting). The ``$d{-}1$ rule'' (use $H_{d-1}$) does not hold simply: $H_2$ features in 3D scale more steeply than $H_1$ features in 2D relative to the shuffled null. The density normalization remains the most effective and most interpretable fix.

\subsection{Berezinskii-Kosterlitz-Thouless transitions}

For the 2D XY model, $\Delta$ peaks at $T \approx 1.0$--$1.2$, well above $T_{\mathrm{BKT}} = 0.8935$. The majority-spin discretization ($\cos(\theta - \bar\theta) > 0$) captures generic ordering, not vortex unbinding. The BKT transition is invisible to this framework. Notably, below-$T_{\mathrm{BKT}}$ data collapse extremely well (collapse quality $1.9 \times 10^{-5}$), suggesting universal scaling in the quasi-ordered phase, but without detecting the transition itself.

\subsection{2D percolation}
\label{sec:percolation}

Site percolation at $p_c$ ($\eta = 5/24$, $d + \eta = 53/24 = 2.208$) provides a test beyond Hamiltonian systems. Using all occupied sites as the point cloud, $\alpha = 1.992 \pm 0.002$, rejected at $> 100\sigma$ from $d + \eta$ and consistent with pure extensivity ($\alpha = d$). This is because occupancy is i.i.d.\ at each site: occupied sites are dense but uncorrelated.

Using only the largest cluster (fractal, $n \sim L^{d_f}$ with $d_f = 91/48$) gives $\Delta < 0$ at all $L$---cluster topology is \emph{suppressed} relative to the shuffled null. The cluster is locally quasi-one-dimensional (a ramified fractal), so its alpha complex has fewer $H_1$ features than a random point cloud of the same density.

The topological gap framework requires a point cloud that is simultaneously \emph{dense} (filling a fraction of the lattice) and \emph{correlated} (spatial arrangement encodes the order parameter). Spin models at $T_c$ provide both via the Fortuin-Kasteleyn cluster structure. Percolation provides either but not both: all sites are dense but uncorrelated; the largest cluster is correlated but sparse and fractal.

\section{Discussion}
\label{sec:discussion}

\begin{table*}[t]
\caption{Summary of all measurements. The scaling law~(\ref{eq:fss})--(\ref{eq:gminus}) is confirmed for 2D second-order transitions and rejected otherwise.}
\label{tab:grand}
\begin{ruledtabular}
\begin{tabular}{lcccccccc}
Model & $d$ & $\eta$ & $d + \eta$ & $\alpha_{\mathrm{meas}} \pm \mathrm{SE}$ & $\sigma_\alpha$ & $\gamma_{\mathrm{MLE}} \pm \mathrm{SE}$ & $1+\beta/\nu$ & Verdict \\
\midrule
2D Ising & 2 & $1/4$ & 2.250 & $2.249 \pm 0.038$ & 0.0 & $1.089 \pm 0.077$ & 1.125 & \textbf{Confirmed} \\
2D Potts $q\!=\!3$$^*$ & 2 & $4/15$ & 2.267 & $2.272 \pm 0.024$ & 0.2 & $1.114 \pm 0.060$ & 1.133 & \textbf{Confirmed} \\
3D Ising$^\dagger$ & 3 & 0.036 & 3.036 & $3.061 \pm 0.044$ & 0.6 & --- & 1.518 & \textbf{Recovered} \\
\midrule
2D Potts $q\!=\!4$ & 2 & $1/2$ & 2.500 & $2.347 \pm 0.017$ & 9.3 & $0.84 \pm 0.60$ & 1.125 & \textbf{Rejected} \\
3D Ising (raw) & 3 & 0.036 & 3.036 & $2.776 \pm 0.065$ & 4.0 & $0.025 \pm 0.003$ & 1.518 & Fails (raw) \\
2D Potts $q\!=\!5$ & 2 & --- & --- & $2.34$ & --- & $4.3 \pm 7.2$ & N/A & Breaks \\
2D XY (BKT) & 2 & $1/4$ & 2.250 & $2.92 \pm 0.12$ & 5.7 & --- & N/A & Not detected \\
2D percolation & 2 & $5/24$ & 2.208 & $1.992 \pm 0.002$ & $>100$ & --- & N/A & \textbf{Rejected} \\
\end{tabular}
\end{ruledtabular}
\flushleft{\footnotesize $^*$$L \geq 64$ fit (5 sizes from $L = 64$--$1024$); all-$L$ fit (6 sizes) gives $\alpha = 2.306 \pm 0.025$ ($1.6\sigma$). Two-term CTS~(\ref{eq:2term_cts}) with $\alpha = d + \eta$ fixed gives $R^2 = 0.9999$ (3 parameters, 6 points).\\$^\dagger$Density-normalized gap $\Delta/|M|^{1/2}$ (Eq.~\ref{eq:delta_norm} with $p = 1/2$).}
\end{table*}

\begin{figure}[t]
\includegraphics[width=\columnwidth]{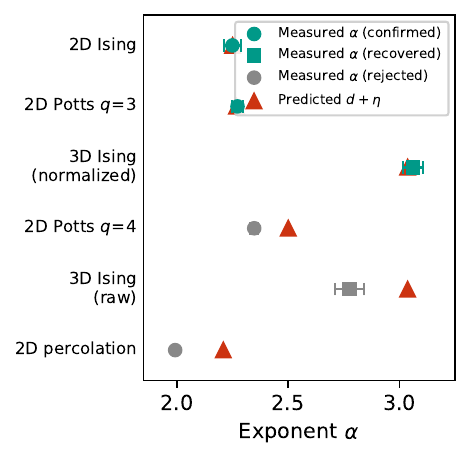}
\caption{Measured $\alpha$ (circles/squares with error bars) vs.\ predicted $d + \eta$ (triangles) for all models. Green: confirmed ($<1\sigma$). Grey: rejected. The 2D Ising and Potts $q = 3$ results match $d + \eta$ to $0.03\sigma$ and $0.2\sigma$ respectively; the $q = 4$ model is rejected at $9.3\sigma$.}
\label{fig:summary}
\end{figure}

The identification $\alpha = d + \eta$ connects persistent homology to the anomalous dimension, a fundamental quantity in the renormalization group. The anomalous dimension $\eta$ governs the power-law decay of the two-point correlation function $\langle \sigma(0)\sigma(r)\rangle \sim r^{-(d-2+\eta)}$ at $T_c$. The per-statistic decomposition suggests a mechanism: at $T_c$, there are $\sim L^{d+\beta/\nu}$ excess $H_1$ cycles in the real versus shuffled data (reflecting fractal cluster structure), and each excess cycle lives $\sim L^{\eta/2}$ longer due to anomalous correlations. In 2D, $2\beta/\nu = \eta$ by hyperscaling, giving $\alpha = (d + \beta/\nu) + \eta/2 = d + \eta$. A companion paper~\cite{loftus2026spectral} provides a more detailed mechanism based on the spectral integral of the connected structure factor.

The scope boundaries have a unifying explanation. The scaling law $\alpha = d + \eta$ requires: (1)~a second-order phase transition with divergent $\xi$; (2)~a majority-spin point cloud that is dense and correlated; (3)~algebraic corrections to scaling ($\omega > 0$). The 2D Ising ($\omega = 2$) and Potts $q = 3$ ($\omega = 4/5$) models satisfy all three conditions. The 3D Ising model satisfies (1) and (3) but requires density normalization to handle the large $\beta/\nu$ dilution. The Potts $q = 4$ model satisfies (1) and (2) but violates~(3): logarithmic corrections ($\omega \to 0$) make convergence unobservable. The remaining failure modes are irreparable:
\begin{itemize}
\item \emph{First-order} ($q = 5$): no divergent $\xi$---correlations are short-ranged.
\item \emph{BKT}: the relevant order is vortex unbinding, not spin alignment.
\item \emph{Percolation}: dense but uncorrelated (i.i.d.), so $\Delta \sim L^d$.
\item \emph{Marginal} ($q = 4$): logarithmic corrections prevent convergence to $d + \eta$ at any accessible $L$.
\end{itemize}

The 2D Potts $G_-$ result ($\gamma = 1.114$, 68\% CI $[1.053, 1.173]$) is consistent with $1 + \beta/\nu$ but cannot distinguish $17/15$ from $9/8$ (0.7\% difference). Resolving this requires either much larger systems or a rigorous derivation of $\gamma = 1 + \beta/\nu$ from first principles.

\section{Conclusion}
\label{sec:conclusion}

We have established a complete finite-size scaling law for the topological gap at criticality: $\Delta(L, T) = A L^{d+\eta} G_-(L|t|)$ with $G_-(x) \sim (1 + x/x_0)^{-(1+\beta/\nu)}$. The exponent $\alpha = d + \eta$ is confirmed for the 2D Ising model ($0.03\sigma$, $L$ up to $1024$) and the 2D Potts $q = 3$ model ($0.2\sigma$, $L$ up to $1024$, with two-term corrections to scaling). The $G_-$ exponent $\gamma = 1 + \beta/\nu$ is confirmed for both classes. In three dimensions, the density-normalized gap $\Delta/|M|^{1/2}$ recovers $\alpha = d + \eta$ to $0.6\sigma$.

The evidence indicates that this scaling law requires algebraic corrections to scaling ($\omega > 0$). The 2D Potts $q = 4$ model---the marginal point with logarithmic corrections ($\omega \to 0$)---is definitively rejected ($9.3\sigma$). The running exponent plateaus at $2.29$ for $L = 128$--$512$, unable to converge to $d + \eta = 5/2$ through the $1/\ln L$ corrections. The framework also fails for first-order transitions, BKT transitions, and percolation.

All results are specific to the alpha complex filtration~\cite{edelsbrunner1994}. Whether $\alpha = d + \eta$ holds for Vietoris-Rips or cubical sublevel filtrations is an open question; if the exponent is a property of the underlying correlated point process rather than the filtration, it should hold generally. Note also that the bootstrap uncertainty on $\gamma$ does not propagate the uncertainty on $\alpha$; we have verified that shifting $\alpha$ by $\pm 0.1$ changes $\gamma_\mathrm{MLE}$ by less than the bootstrap SE, so this is a subdominant effect.

Open problems include: (i)~a rigorous derivation of $\alpha = d + \eta$ and $\gamma = 1 + \beta/\nu$ from the statistical mechanics of alpha complexes on correlated point processes (a companion paper~\cite{loftus2026spectral} provides a partial answer through the spectral integral of the connected structure factor); (ii)~deriving the modified scaling function $\tilde{G}_-$ in the density-normalized 3D case and identifying its exponent $\gamma \approx 0.25$ in terms of critical exponents; and (iii)~testing $\alpha = d + \eta$ in four or more dimensions, where the density dilution is even more severe and higher-order density normalization may be required.

\begin{acknowledgments}
Computations used the GUDHI library~\cite{maria2014} for persistent homology and SciPy for Swendsen-Wang cluster decomposition.
\end{acknowledgments}

\end{document}